\documentclass[a4]{article}

\usepackage{amsmath}
\usepackage{graphicx}
\usepackage{bbm}
\usepackage{color}

\usepackage{cite}

\newcommand{\bra}[1]{\langle #1 |} 
\newcommand{\ket}[1]{| #1 \rangle } 

\definecolor{cbl}{rgb}{0,0,1}

\definecolor{crd}{rgb}{1,0,0}

\title{Controlling quantum flux through measurement:\\ an idealised example}
\author{Antoine Tilloy ${}^{\clubsuit~}$, Michel Bauer ${}^{\spadesuit~}$ and Denis Bernard ${}^{\clubsuit~}$}

\begin{document}

\maketitle

\vskip -0.5 truecm 

\begin{center}
{\footnotesize 
\noindent
${}^\spadesuit$ Institut de Physique Th\'eorique de Saclay, CEA-Saclay $\&$ CNRS, 91191 Gif-sur-Yvette, France.\\
$^\clubsuit$ Laboratoire de Physique Th\'eorique de l'ENS, CNRS $\&$ Ecole Normale Sup\'erieure de Paris, France.}
\end{center}
\vskip 0.5 truecm

\begin{abstract}
Classically, no transfer occurs between two equally filled reservoirs no matter how one looks at them, but the situation can be different quantum mechanically. This paradoxically surprising phenomenon rests on the distinctive property of the quantum world that one cannot stare at a system without disturbing it. It was recently discovered that this seemingly annoying feature could be harnessed to control small quantum systems using weak measurements. Here we present one of the simplest models -- an idealised double quantum dot -- where by toying with the dot measurement strength, i.e. the intensity of the look, it is possible to create a particle flux in an otherwise completely symmetric system. The basic property underlying this phenomena is that measurement disturbances are very different on a system evolving unitarily and a system evolving dissipatively. This effect shows that adaptive measurements can have dramatic effects enabling transport control but possibly inducing biases in the measurement of macroscopic quantities if not handled with care.

\end{abstract}

\section{Introduction}

Monitoring a quantum system  -- i.e. observing it repeatedly and frequently -- induces stochastic evolutions, or \emph{quantum trajectories}\cite{carmichael1993,breuer2002,wiseman1996}, where the uncertainty of the outcomes is the source of the randomness. Quantum jumps\cite{bohr}, which consist in the sharp transitions of a system from one quantum state to another,  are fundamental illustrations of this behaviour. They have been observed for the first time in fluorescent systems\cite{nagourney,sauter} and nowadays in almost everyday meso- or nano-scale experiments\cite{vijay2012,ExpFluct,murch2013}. They emerge naturally as a consequence of the interplay between system evolution and continuous measurement and are very abrupt yet not strictly instantaneous.  Aside from their use in quantum Monte Carlo methods\cite{dalibard1992}, quantum trajectories are instrumental to control small quantum systems\cite{wiseman2010,thomsen2002,ruskov2002}. Feedback control consists in using the measurement record to act on the system in order to constrain or control its evolution. The action on the system can be carried out through a Hamiltonian drive -- the most common and historical method\cite{ahn2002,sarovar2004} -- yet it was underlined recently\cite{erez2008,jacobs2010,blok2014} that measurements themselves could be used as they inevitably have a back-action on the system. A minimal version of this second method deals with adjusting the measurement intensity (or rate) but not the measured observables. Quantum Zeno dynamics\cite{facchi2000,signoles}, in which the dynamics of the system is frozen in a subspace selected through strong measurement, is an archetypal example. However control by measurement is multi-faceted because the latter picture is true only for a system evolving unitarily and breaks down for a system evolving dissipatively. Indeed, Hamiltonian and thermally-activated quantum jumps behave very differently with respect to a varying measurement intensity. In the first case, the Zeno effect makes the average time between two jumps increase with the measurement strength whereas in the dissipative case it 
converges to 
a constant which is intrinsic to the system-reservoir coupling\footnote{The main physical rationale behind this fact lies on the very large number of degrees of freedom of the reservoir dissipatively coupled to the system. }. 
In this paper we consider a model system, idealising a double quantum dot in contact with two electron reservoirs, where both a unitary and a dissipative evolution compete.This is the simplest system where the effect we are ultimately interested in can be seen. The tunnelling rate {between} dots can be controlled by adjusting the intensity of the particle position measurement without impacting the dot-reservoir jump rates. Available pathways can thus be controlled through measurement, and this is the basic principle making flux control possible. By monitoring the particle position, we may choose to measure strongly when the particle is in the right dot and mildly otherwise. As a consequence, when the particle is in the right dot, the jump rate to the right reservoir is not modified but its tunnelling rate to the left dot is frozen and the probability to backtrack is thereby reduced. This creates a {non-zero} extra average particle flux from the left reservoir to the right, and a macroscopic effect is obtained from what seemed to be a very innocent measurement method. 

\section{A simple model}

\begin{figure}
\centering
\includegraphics[width=0.14\textwidth]{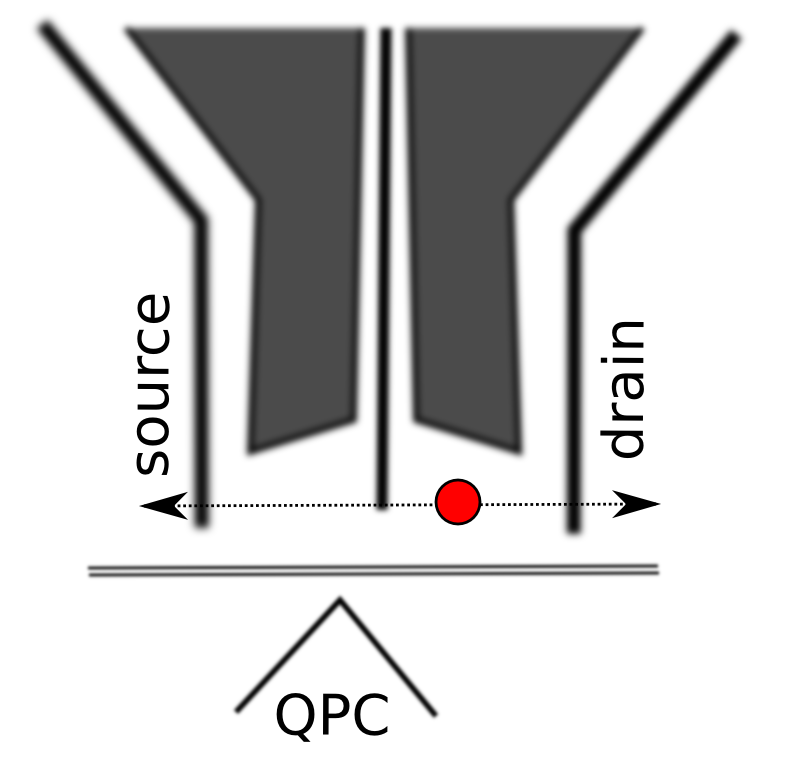}
\includegraphics[width=0.34\textwidth]{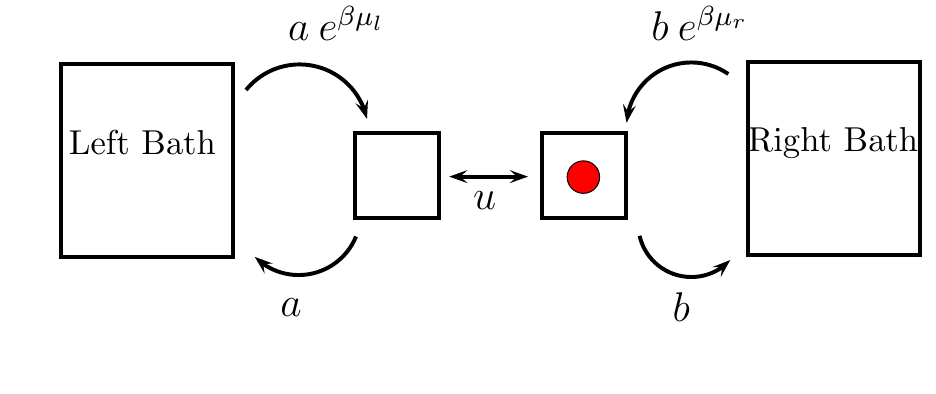}
\caption{Schematics of a double quantum dot (DQD) measured by a quantum point contact (QPC) and the idealised model we derive from it, with an electron on the right. The arrows represent the possible dynamical processes.}
\label{fig:ideal}
\end{figure}

The physical system we have in mind is that of a double quantum dot (DQD), or rather an abstract idealisation of a DQD\footnote{The latter should be thought of as the inspiration for our model rather than its justification as we obviously neglect many important features of real DQDs. Circuit QED systems may actually be better candidates for a potential experimental realisation.} with an architecture similar to that of Fig. \ref{fig:ideal}, coupled to two electron reservoirs at given chemical potentials. We may schematically picture our model system as in Fig. \ref{fig:ideal}, in which electrons can either {jump to} the left/right dots from the left/right reservoirs, back and forth, or tunnel through the potential barrier from one dot to the other. The position of the electron is continuously monitored with a non-specified measurement apparatus, which in the DQD architecture can be implemented through an auxiliary quantum point contact (QPC) as in \cite{ExpFluct,Milburn2001,EffFluct}. We further assume that the 
Coulomb repulsion between two electrons in the quantum dots is high enough to be able to neglect double 
occupancy in our idealised description. We write $\ket{0}$, $\ket{L}$, and $\ket{R}$, the system space vectors where there is no electron in the DQD, one in the left dot, and one in the right dot, respectively. Three processes are at play: tunnelling {between} the dots; quantum {hopping} {between} dots and reservoirs; and back-action of the measurement. They generate the DQD density matrix evolution, called the quantum trajectory:
\begin{equation}
\label{eq:evol}
d\rho_t=d\rho_t^{tunnel}+d\rho_t^{bath}+d\rho_t^{measure},\\
\end{equation}
during a time duration $dt$. The {tunnelling} evolution is unitary, $d\rho_t^{tunnel}=-i[H,\rho_t] dt$ with inter-dots Hamiltonian,
\begin{equation}
 H= u \left(\ket{R}\bra{L}+\ket{L}\bra{R}\right).
\end{equation}
The dot-bath {hopping} generated evolution $d\rho_t^{bath}$ is dissipative, and we model it through a deterministic evolution of the form $d\rho_t^{bath}= \mathcal{L}^{bath}(\rho_t)\, dt$ with $\mathcal{L}^{bath}(\cdot)$ a Lindblad operator\cite{lindblad1976} linear in $\rho_t$. We assume that, in the absence of tunnelling, the left reservoir would thermalise the left dot into a Gibbs steady state of chemical potential $\mu_l$ (resp. $\mu_r$ for the right reservoir and right dot). We denote by $a$ (resp. $ae^{\beta\mu_l}$) the {hopping} rate from the left dot to the left reservoir (resp. from the left reservoir to the left dot) in absence of tunnelling, and by $b$ and $be^{\beta\mu_r}$ the respective right dot-reservoir {hopping} rates {($\beta$ being the inverse temperature)}.  A possible implementation of this prescription is to write $\mathcal{L}^{bath}(\rho_t)=\big( L_{\sigma^{+}_{l}}+L_{\sigma^{-}_{l}}+L_{\sigma^{+}_{r}}+L_{\sigma^{-}_{r}}\big) (\rho_t) $ where the associated Lindblad generators $L_{\sigma^{+}_{l}}$, $L_{\sigma^{-}_{l}}$, $L_{\sigma^{+}_{r}}
$, $L_{\sigma^{-}_{r}}$ take the form $L_\sigma(\rho)=\sigma\rho \sigma^\dagger - \frac{1}{2}\lbrace \sigma^\dagger \sigma, \rho\rbrace $ with $\sigma^{+}_{l}=\sqrt{a} e^{\beta \mu_l/2} \ket{L} \bra{0}$, $\sigma^{-}_{l} = \sqrt{a} \ket{0} \bra{L}$, $\sigma^{+}_{r}= \sqrt{b} e^{\beta\mu_r/2} \ket{R} \bra{0}$ and $\sigma^{-}_{r} = \sqrt{b} \ket{0} \bra{R}$.
The left/right dot occupancy is monitored continuously via a weak measurement apparatus. We write $\mathcal{O}=h_l \ket{L}\bra{L} + h_r \ket{R} \bra{R}+h_0 \ket{0}\bra{0}$ the measurement operator
(in the following, we take $h_0=0$ because only the differences matter), the strength of the measurement, i.e. the intensity of the {\it look}, increases with the values of $h_{r,l}$.  The measurement result is a random process, which we denote $X_t$. Physically, $X_t$ may represent the total charge going through the quantum point contact monitoring the DQD, the state being read from the local average intensity (the instantaneous intensity being, strictly speaking, singular) as was done in \cite{ExpFluct}. 
It is known\cite{wiseman2010,barchielli1986,barchielli1991,belavkin1992} that quantum mechanical rules for measurement imply that $X_t$ varies in time according to: $dX_t= 2\, \mathrm{tr}(\mathcal{O}\rho_t) dt + dW_t$. Here, $W_t$ is a standard Wiener process, i.e. ${\mathrm{d}W_t}/{\mathrm{d}t}$ is the usual white noise in-time process, and its randomness is an echo of the random nature of quantum measurements. A given realisation of this Wiener process corresponds to a given realisation of a time-series of weak measurements.
It is also known\cite{wiseman2010,barchielli1986,barchielli1991,belavkin1992}  that the back-action of the measurement on the system density matrix reads $d\rho_t^{measure}=L_{\mathcal{O}}(\rho_t) \, dt+ D_{\mathcal{O}} (\rho_t) dW_t$ {where $L_{\mathcal{O}}(\cdot)$ is the Lindblad generator associated to $\mathcal{O}$ and} where $D_{\mathcal{O}} (\cdot)$ is the stochastic innovation term, $D_{\mathcal{O}} (\rho)= \left\lbrace \mathcal{O},\rho \right\rbrace-2\, \rho \, \mathrm{tr}(\mathcal{O}\rho)$. It is worth noticing that any given realisation of a time-series of weak measurement, i.e. a time-series of observation, corresponds to a realisation of the process $X_t$, and hence that the process $X_t$ is the only information our model observer possesses on the DQD.

As a further simplification we take the DQD density matrix $\rho_t$ to be of the following form, 
\begin{equation}
\label{eq:rho}
\rho=\left[
\begin{array}{ccc}
Q_0 &0&0\\
0& Q_l&iK/2\\
0& -iK/2&Q_r
\end{array} \right].
\end{equation}
This choice {can actually be made with no lack of generality as a diagonal density matrix is naturally prepared and as the form (\ref{eq:rho}) is preserved by the evolution (\ref{eq:evol})}. This parametrisation of the density matrix simply shows the probabilities $Q_0$, $Q_l$ and $Q_r$ for the DQD to be in the basis states $\ket{0}$, $\ket{L}$, and $\ket{R}$, plus a (real) phase $K$ encoding the delocalisation due to the tunnel coupling between the dots. We now have everything needed to expand eq. (\ref{eq:evol}) and get explicit evolution equations for $Q_0$, $Q_r$, $Q_r$ and $K$, see eq.~ \eqref{eq:Expanded}. {We have used the compact notation $m_l=e^{\beta \mu_l}$, $m_r=e^{\beta \mu_r}$ and $\nu_h=h^2+(a+b)/4$.}
\begin{equation}
 \label{eq:Expanded}
 \left\lbrace
 \begin{split}
  dQ_0&=\big(a Q_l+b Q_r-(a m_l+b m_r) Q_0\big)\, dt -2 \big(h_l Q_l + h_r Q_r\big) Q_0\, dW_t\\
  dQ_l&=\big(-uK+a m_l Q_0-a Q_l\big) \, dt  +  2\left(h_l(1-Q_l )-h_r Q_r\right) Q_l \,dW_t  \\
  dQ_r&=\big(uK-b Q_r+b m_r Q_0\big) \, dt + 2\left(h_r(1-Q_r)-h_l Q_l\right) Q_r \, dW_t \\
  dK&=2\big(-\nu_h\, K + u(Q_l-Q_r)\big) \, dt + (h_l(1-2Q_l)+h_r(1-2Q_r))K \: dW_t
 \end{split} \right.
\end{equation}

\section{Quantum jumps}

\begin{figure}
\centering
\includegraphics[width=0.46\textwidth]{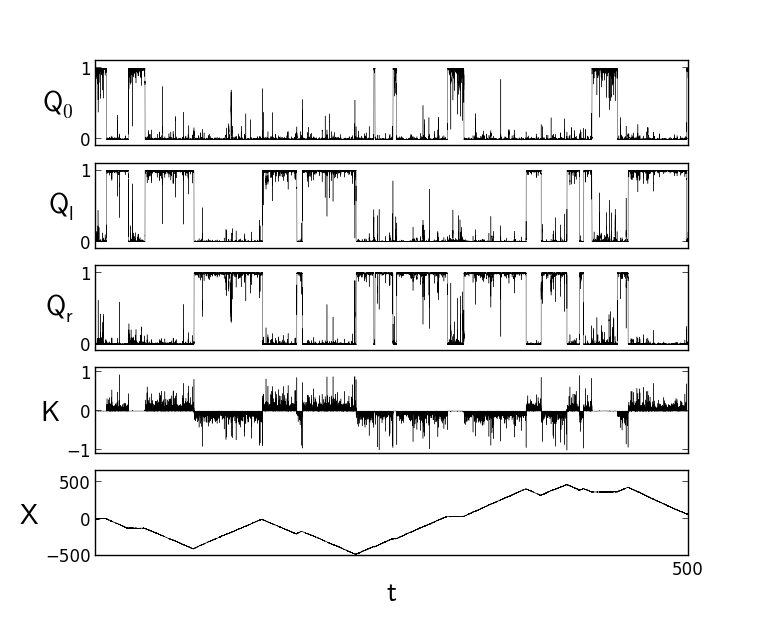}
\caption{A sample trajectory with $a=b=0.02$, $\beta \mu_l=-\beta \mu_r =1.0$, $h_r=-h_l=7.0$ and $u=1.0$.}
\label{fig:trajectory}
\end{figure}

Equation (\ref{eq:evol}) is a set of stochastic differential equations (SDEs) for the DQD density matrix coefficients. It is very complicated, there is no hope to solve it exactly, and the description may look a bit complicated at this point. However it is obvious from Fig.\ref{fig:trajectory} that the behaviour of the system is very peculiar in the large $h_l \, \& \, h_r$ limit. Indeed, if the measurement strength is strong enough the system has a tendency to collapse most of the time {onto} one of the three states $\ket{0}$, $\ket{L}$, and $\ket{R}$, and do random jumps between them. These are the quantum jumps, and we can \emph{approximately} describe the DQD evolution  as a time-series of random jumps between those three states.
This limiting description, which is often assumed from the start and which is admittedly much simpler but only valid in an 
appropriate regime, is actually not trivial to derive from the full description we have provided. Proving rigorously that it does so will not be done in this paper, rather we will invoke heuristic arguments and numerical simulations.
Because $\mathrm{tr}(\rho)=1$, the diagonal coefficients of the density matrix evolve inside a triangle. When the monitoring is not strong enough, their random trajectory explores the whole bulk but when it is intense, they spend most of their time near the corners. From a physical point of view {this is completely} expected: indirectly measuring the particle position has a tendency to make the wave packet collapse {onto} one of the position eigenstates.

\begin{figure}
\centering
\includegraphics[width=0.25\textwidth]{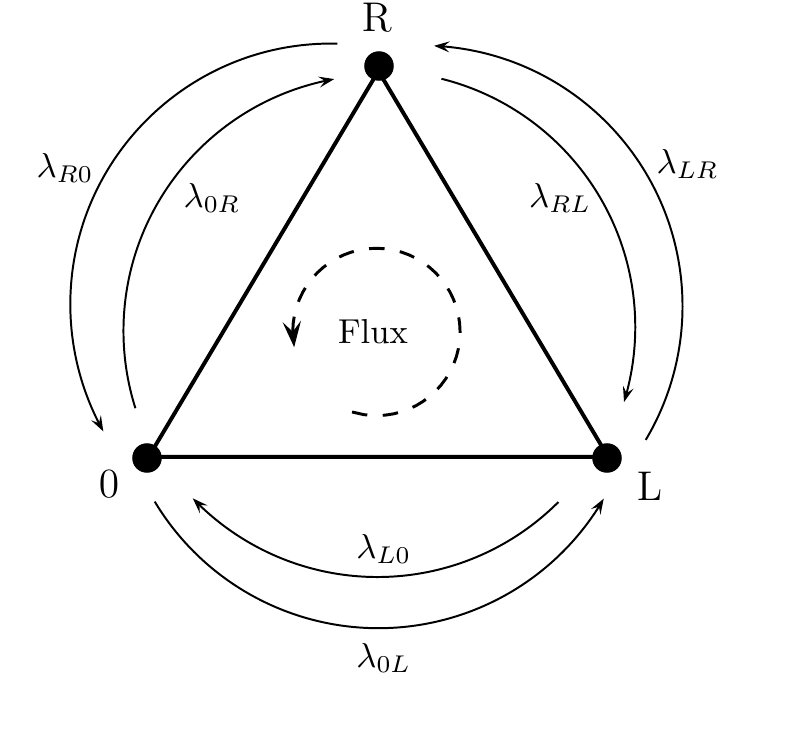}
\caption{Approximation of the evolution as a Markov chain on a simplex.}
\label{fig:markov}
\end{figure}

We thus have an approximated effective Markov chain description of our system in the strong measurement limit (see Fig. \ref{fig:markov}), in which the DQD system is described by the probabilities $\Pi_{0,l,r}$ for the DQD density matrix to be near one of those triangle vertices, that is, the probabilities for the DQD state to be close to $\ket{0}$, $\ket{L}$, and $\ket{R}$. The time evolution of these probabilities is fully characterised by the transition rates between the different states. Their direct computation from eq. \eqref{eq:Expanded} is technical, and will be done properly in \cite{inpreparation}. {As suggested by the numerics, let us assume} that the $0\leftrightarrow L$ transition process is dominated by trajectories close to the $0L$ boundary of the simplex for which $Q_r\simeq 0$ (and hence $K\simeq0$ and $Q_l \simeq 1-Q_0$). Under these approximations, eq.(\ref{eq:Expanded}) then becomes $dQ_0=\lambda (p-Q_0) dt +2 h (1-Q_0) Q_0\, dW_t$ with $\lambda=a(m_l+1)+b m_r$ and $p=a/\lambda$. This evolution {encodes} the competition between the bath coupling and the measurement back-action, see e.g. \cite{thermalImaging}. The sharp transitions can be understood by analogy with Kramers' theory where a particle excited by white noise fluctuations shows a jumpy behaviour because it evolves in a bimodal effective potential. The jump statistics from $Q_0=0$ to $Q_0=1$ (or the reversed) becomes Poissonian in the large $h$ limit, the transition rates, defined as the mean time between two jumps, were shown to converge to $\lambda_{0L}=am_l$, $\lambda_{L0}=a$ (and similarly $\lambda_{0R}=bm_r$, $\lambda_{R0}=b$) for $h$ large. {Similar methods along the lines of \cite{oqrw,qbm} enable one to estimate} the transition rates in the $R\leftrightarrow L$ region. {However, these arguments assume a decoupling between the different processes which has to be proved \cite{in preparation}.} In what follows, we will only use the fact (see \cite{inpreparation}) that in the strong measurement 
limit, these transition rates are: $\lambda_{LR}=\lambda_{RL}\simeq u^2/h^2$, $\lambda_{0L}\simeq a m_l$, $\lambda_{L0}\simeq a$, $\lambda_{0R}\simeq bm_r$ and $\lambda_{R0}\simeq b$. The point to notice is that the rates $\lambda_{LR}$ and $\lambda_{RL}$, induced by the tunnelling transition $R \leftrightarrow L$, vanish as $h^{-2}$ for strong measurement in accordance 
with the quantum Zeno effect, while the rates $\lambda_{0R}$ ($\lambda_{0L}$) and $\lambda_{R0}$ ($\lambda_{L0}$), activated by the dissipative contacts $0\leftrightarrow R(L)$, are not renormalised {neither by} the tunnelling nor the measurement and remain finite in the strong measurement limit.

\section{Measuring the flux without feedback}

\begin{figure}
\centering
\includegraphics[width=0.48\textwidth]{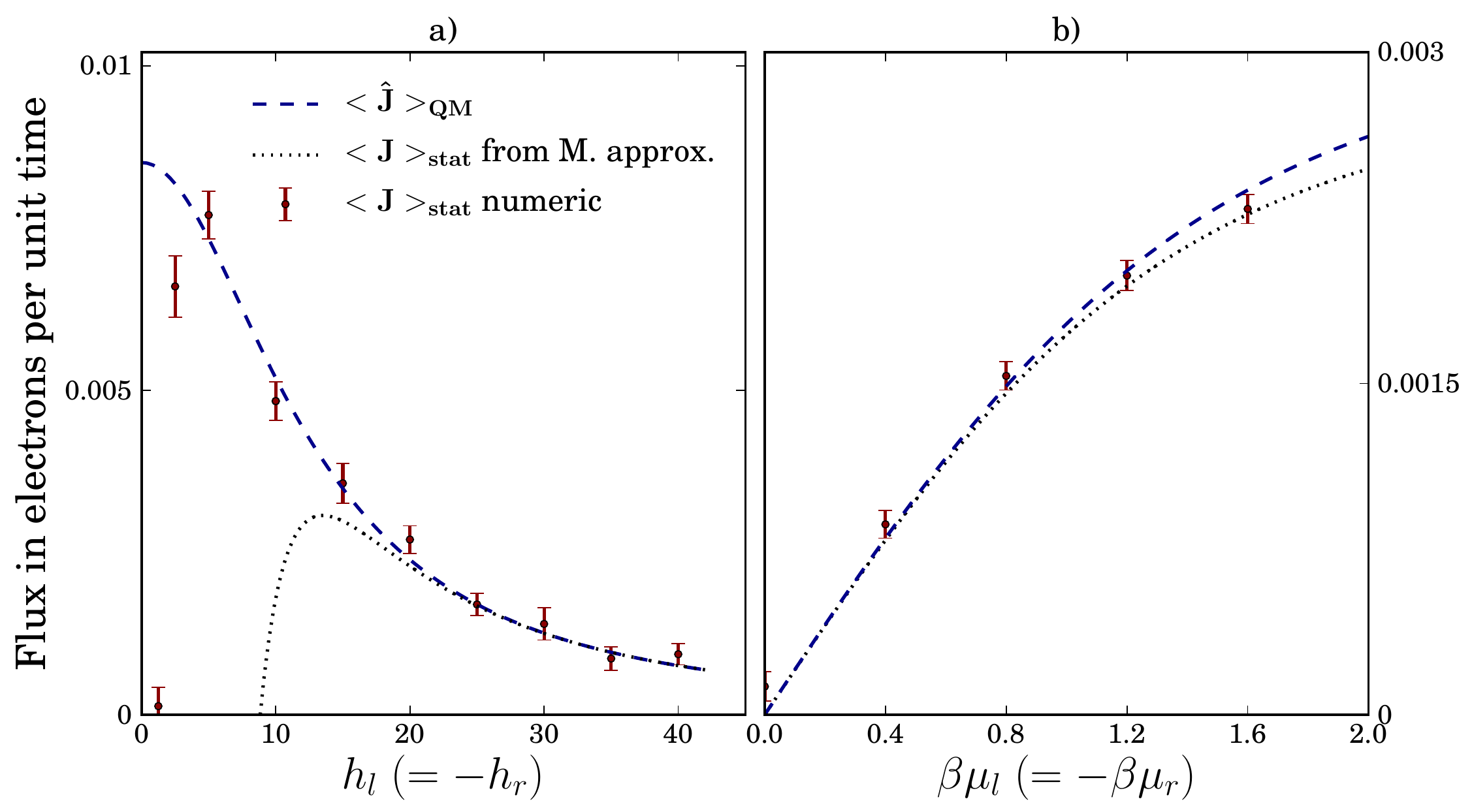}
\caption{Comparison between purely analytic {(quantum mechanical average)} flux, Markov approximated and numeric {statistical} flux as a function of the measurement strength with $a=b=0.02$ and $u=1.0$ 
  \textbf{a) } Constant difference of potential $\beta\mu_l=-\beta\mu_r=2.0$ and the measurement strength varies. Notice that the three methods converge for strong measurements as expected.
  \textbf{b) } Constant measurement strength $h_l=-h_r=15.0$ and the difference of potential varies.}
\label{fig:flux}
\end{figure}

By monitoring the particle position continuously in time we have access to the electron flux. However the only information our model observer has on the DQD is the measurement output $X_t$, and thus she/he has to read out the flux from this process\footnote{We are not adding another flux measurement device that we would have to describe a la Von Neumann.}. Since for strong enough measurement, the particle position is {approximately} well defined,  it is possible to rely on a classically inspired characterisation of the flux. One can simply define the electron transfer during a time duration $t$ as the difference between the number $N_{LR}(t)$ of quantum jumps from state $\ket{L}$ to state $\ket{R}$ and the number $N_{RL}(t)$ of quantum jumps from $\ket{R}$ to $\ket{L}$ during this time period. This can be read out from the numbers of break points in the time-slope of $X_t$, see Fig.\ref{fig:trajectory}. This is a random number whose statistics is induced from that of the DQD density matrix 
components. Hence, the mean flux that the observer is measuring is the ``statistical mean" of the electron transfer (and not a quantum average in the usual sense, see below).  By ergodicity, averaging in time yields the mean, and the ``statistically computed" mean flux is naturally defined as \footnote{This definition only makes sense for strong enough measurement as otherwise quantum jumps are not well defined: they become fuzzier as the measurement becomes weaker. We also assume that the time sampling is refined enough (the detector bandwidth large enough) to resolve {fast chains} of transitions, e.g to resolve the chain of events $L\to 0\to R$ from the event $L\to R$ .}:
\begin{equation} \label{eq:StatJ}
\langle J\rangle_{stat} = \lim_{t\rightarrow +\infty} \frac{1}{t} \left({N_{LR}(t)-N_{RL}(t)}\right) .
\end{equation}
We numerically solved the system of SDEs (\ref{eq:evol}) for the DQD quantum trajectory -- and not the approximate Markov chain -- and computed the measured average flux we described in (\ref{eq:StatJ}). {As} the equations are strongly non linear, we used an adaptive Runge-Kutta discretisation method  which is described in more details in \cite{inpreparation}. The results are shown in Fig. (\ref{fig:flux}). The flux decreases as a function of the measurement strength: {strong measurements tend to Zeno freeze the tunnelling transition between $\ket{L}$ and $\ket{R}$ thus making it harder for the electrons to go through the DQD.}

Within the Markov chain approximation this mean flux can be easily computed. Standard results from probability theory \cite{jacod2003} allow {the flux to be expressed} as $\lambda_{LR} \Pi_l^{stat}- \lambda_{RL} \Pi_r^{stat}$ with $\Pi_l^{stat}$ (resp.  $\Pi_r^{stat}$) the Markov chain stationary probability to be in $L$ (resp. $R$). The stationary probabilities can be easily computed from the Markov master equation implied by the transitions rates (see \cite{in preparation} for more details) which gives:
\begin{equation}
\begin{split}
\langle J\rangle_{stat} &= \frac{(m_l-m_r)  u^2}{\nu_h (m_l+m_r+1)+u^2\left(\frac{1+2m_l}{b}+\frac{1+2m_r}{a}\right)}\\
&\simeq \frac{(m_l-m_r)  u^2}{h^2 (m_l+m_r+1)}
 \end{split}
\end{equation}
in the strong measurement limit and with $h_l-h_r=2h$. This expression is plotted in Fig.\ref{fig:flux} for comparison with the exact numerical result, and clearly confirms the accuracy of our effective description in the strong measurement limit.

How {would one} compute the electron flux in the standard way without taking into account the measurement records? A reasonable electron flux operator in our setting is the following:
\begin{equation}
\hat J = i \, u \left( \ket{R} \bra{L}-\ket{L} \bra{R}\right)
\label{eq:fluxOp}
\end{equation}
Quantum mechanically, the average flux is then obtained by taking the trace of this operator against the steady DQD density matrix: $\langle \hat J\rangle_{MQ}=\mathrm{tr}(J \bar \rho^{steady}) =u \bar K^{steady}$. One has to pick the mean DQD density matrix because one is not recording the measurement outputs (and taking the mean means averaging over all possible DQD quantum trajectories). As expected and shown in Fig.\ref{fig:flux}, $\langle \hat J\rangle_{MQ}=\langle J\rangle_{stat}$, but for this to hold one has to include the back-action effect of the continuous monitoring on the DQD density matrix (even though we did not record the measurement results). 

This result is admittedly expected, but the way it emerges from completely different descriptions and equations is nevertheless surprising, and as far as we know, this connection has never been made explicitly before.
The statistical definition for the flux has the advantage {of being} usable also in the case where there is no steady density matrix, for example in the presence of a general feedback scheme.

\section{Feedback}

The peculiarity of the present situation which allows for feedback is that our model observer gets information on the flux continuously in time and realisation by realisation, i.e. not as an ensemble average. The main idea behind the feedback rests on the fact that the transition rates for the Hamiltonian junction or the dissipative coupling do not scale in the same way as the measurement strength increases. By measuring more strongly when the particle is on the right we increase its probability to jump in the right reservoir rather than backtrack to the left dot. If we continuously monitor the particle position we can increase the measurement intensity only when it is on the right thus breaking the symmetry of the system and creating a flux from left to right. 

\begin{figure}
\centering
\includegraphics[width=0.49\textwidth]{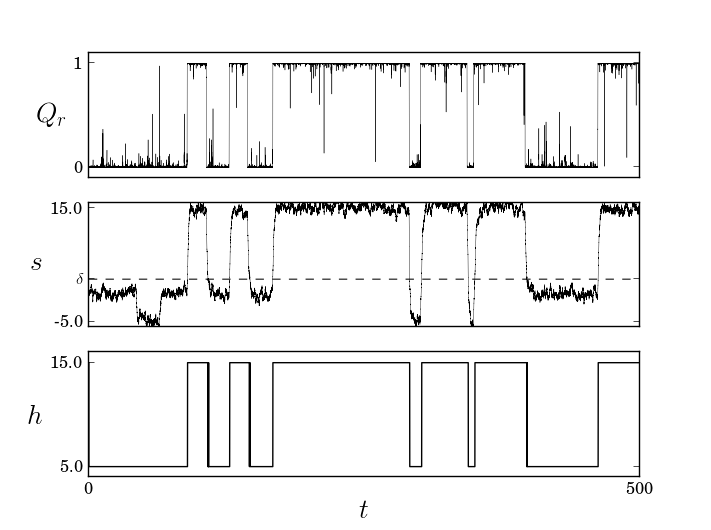}
\caption{A snapshot of a trajectory with measurement feedback $a=b=0.02$, $\beta \mu_l=\beta \mu_r =0.0$, $u=1.0$, $\delta=2.0$, $\tau_{int}=1$, $h_{min}=5.0$ and $h_{max}=15.0$
}
\label{fig:feedback}
\end{figure}

In order to control the system, the first thing to do is to infer the value of the density matrix $\rho_t$ from the measurement records $X_t$. Since the time  evolution of the DQD is {formally} driven{\footnote{Indeed, $d\rho_t=\left[\mathcal{L}_{tot}(\rho_t)-2tr(\mathcal{O}\rho)D_{\mathcal{O}}(\rho_t)\right] dt +D_{\mathcal{O}}(\rho_t) dX_t$ where the deterministic part of $\rho$ is collected in $\mathcal{L}_{tot}(\rho_t)$.}} by $X_t$, one could theoretically extract exactly $\rho_t$ from $X_t$. However it is more practical\cite{sarovar2004} and more robust to  estimate it using the fact that $\rho_t$ is given by the slope of the measurement records in the strong measurement limit. 
Indeed, the drift term in the time evolution of $X_t$, i.e. $dX_t= 2\, \mathrm{tr}(\mathcal{O}\rho_t) dt + dW_t$, overcomes the noise for strong measurement. It is equal to $0$ when $Q_0\simeq 1$, $2 h_l$ when $Q_l\simeq 1$ and $2h_r$ when $Q_r\simeq 1$. So when the density matrix is concentrated on one of the natural basis states and if $h_l$ and $h_r$ are big enough, i.e. the measurement is strong enough, the DQD density matrix $\rho_t$ can be approximately read out from the time-slope of the measurement results. 
Besides, this instantaneous slope is (formally) infinite because of white noise properties, and it consequently needs to be averaged over a small time window $\tau_{int}$. We thus write $s_t=\int_{-\infty}^{t} e^{(t-\tau)/\tau_{int}} dX_t$. Further we take a measurement operator that discriminates equally L and R from 0, i.e, we choose $h_r=-h_l=h > 0$. We then measure {as strongly} as we can when the electron is likely to be in the right dot and keep measuring mildly when it is not to keep track of its position. This can be enforced {by} taking: 
$h(t)= h_{min} \, \mathbf{1}_{s_t \leq \delta} + h_{max}\, \mathbf{1}_{s_t > \delta}$, 
where $\delta$ is an arbitrary threshold that can be optimised numerically {and $\mathbf{1_{x<y}}$ is the indicator function, equal to $1$ if the inequality is satisfied and $0$ otherwise}. A snapshot of a trajectory is given in Fig. \ref{fig:feedback}. The evolution of the system now obeys an even more complicated stochastic differential equation. However, assuming that the measurement stays strong enough for the Markovian approximation to be faithful, we can estimate the maximum flux that would be obtained with a perfect feedback by simply changing the jump rates, i.e. computing $\lambda_{LR}$, $\lambda_{0L}$, $\lambda_{0R}$, $\lambda_{L0}$ with $h=h_{min}$ and $\lambda_{RL}$, $\lambda_{R0}$ with $h=h_{max}$. Actually, only  $\lambda_{RL}$ and $\lambda_{LR}$ depend on $h$ so we only  have to change two rates. This gives:
\begin{equation}
\langle J\rangle_{stat} \simeq \frac{u^2}{1+m_l+m_r} \left[ \frac{m_l}{h_{min}^2}-\frac{m_r}{h_{max}^2}\right]
\end{equation}
That is, as expected, a {non-zero} flux even when the two chemical potentials are equal. This result is valid only for a perfect feedback scheme, that is if the system state is perfectly known at all {times}. As a result it is unfortunately just an upper bound for practical feedback schemes. However, this feedback scheme is robust enough for the existence of an induced current to persist away from the strong measurement regime.

\begin{figure}
\centering
\includegraphics[width=0.48\textwidth]{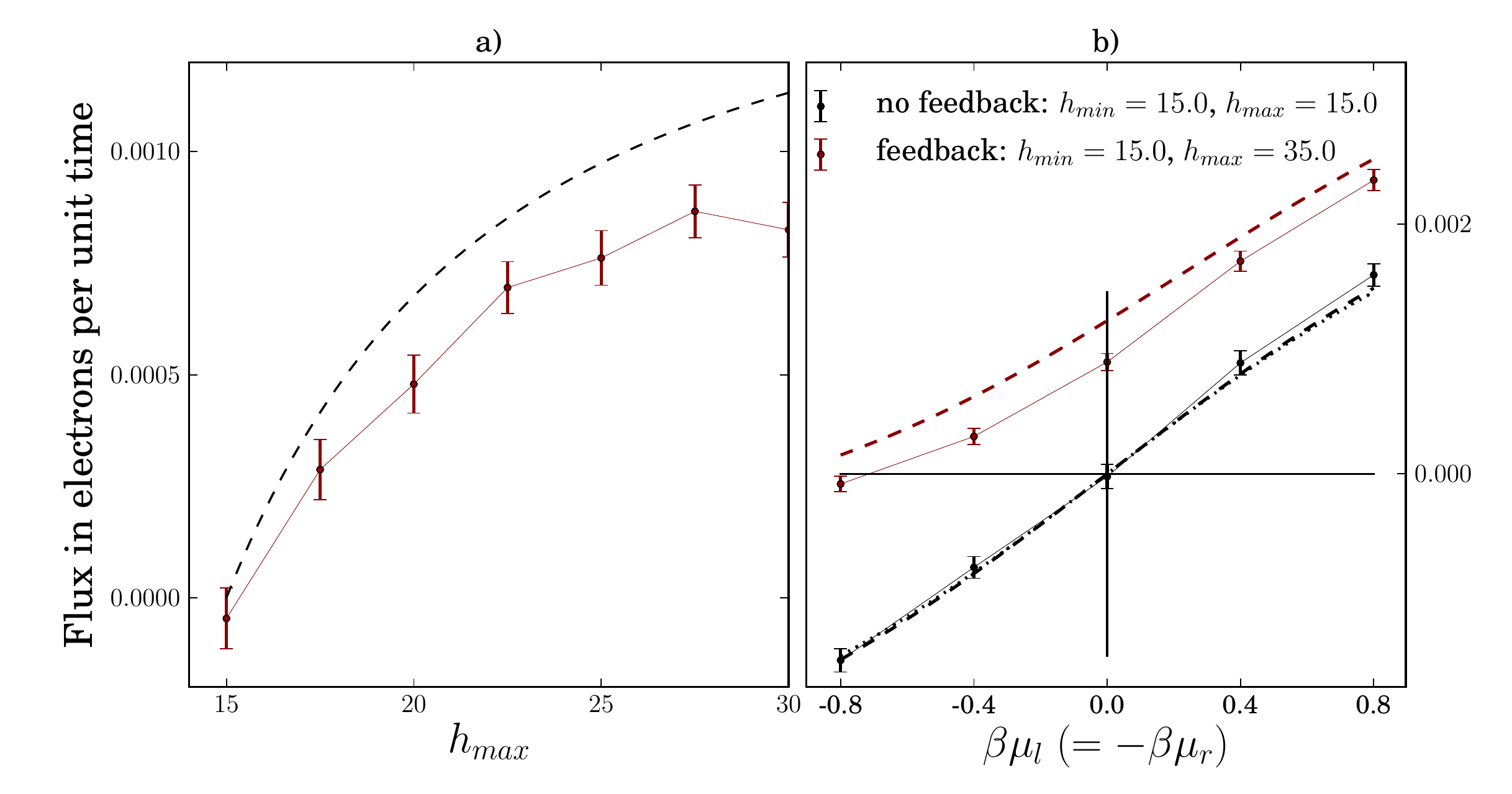}
\caption{\textbf{Flux with feedback}
\textbf{a)} Average electron flux as a function of the feedback strength $h_{max}$. The analytic upper bound previously found is shown {by a} dashed line. Here $h_{min}=15$, $a=b=0.02$, $u=1.0$,$\delta=4.0$, $\tau_{int}=0.5$ and $\mu_{l}=\mu_r=0$ that is no potential difference between the two baths. The feedback scheme creates the flux in the DQD as expected.
\textbf{b)} Average electron flux as a function of the difference of chemical potential between the two electron bath. The analytic expressions in the Markovian limit are shown {by} dashed lines. The feedback scheme indeed increases the flux in the DQD and is even able to counter a small difference of potential. ($a=b=0.02$, $u=1.0$, $\delta=2.0$, $\tau_{int}=1$)}
\label{fig:feedbackFlux}
\end{figure}

Using the same numerical method as before, we estimate the electron flux going through the idealised DQD for various feedback strengths, i.e. for various maximum measurement rates. We compared the numerical flux with the upper bounds we have just found. The numerical computation were done with complete SDEs, including feedback, and not with the Markovian approximation. The results are shown on Fig.\ref{fig:feedbackFlux}. In particular, they clearly show that this feedback procedure is able to generate a flux in {the} absence of a chemical potential difference. {The disagreement between the upper bound and the practical feedback scheme for strong feedback is a consequence of the finite window of integration of the measurement records: as the jumps get sharper it becomes increasingly difficult to know the system state in time and to switch the measurement strength fast enough.}

\section{Discussion}

Using a simple feedback scheme, in the sense that it relies on the measurement {strength of the} apparatus only, we have shown that it is possible to create, control or reverse a particle flux in a {quantum system}. {Even if our study has been done in a simple setting, the effect we have discovered is undoubtedly very general. The way we monitor a quantum system not only modifies its perceived fluctuations or correlations but also changes its non-equilibrium properties in a dramatic way (changing the sign of a flux in our example). This remains true even if the measurement itself has a completely symmetric back-action on the whole system.}
In this respect, the effect we described bears similarities with the Parrondo paradox\cite{harmer2002} or flashing ratchets \cite{makhnovskii2004,lopez2008,craig2008} because it amounts to {the production of} an asymmetric output by oscillating between two different yet symmetric situations.
{This effect could be used to engineer out-of-equilibrium quantities in systems where the measurement intensity is the only control parameter. It needs to be taken into account in the numerous experiments that use adaptive measurement schemes to measure fluxes or other macroscopic quantities as it could introduce substantial biases in the results.}

\paragraph*{Acknowledgements}
The authors would like to thank A. de Luca and J. Viti for useful discussions, and B. Huard for his experimental insights. This work was in part supported by ANR contract ANR-2010-BLANC-0414.
 

 \vfil \eject
 
\appendix

\noindent {\bf \LARGE Supplementary Materials:}
\bigskip

\section{Analytics}

We here give details on the behaviours of the  DQD quantum trajectories and discuss how they can approximatively be described by a three state Markov chain.

\subsection{Structural properties of the DQD quantum trajectories}

Time evolution of the DQD quantum trajectory is governed by $d\rho_t=d\rho_t^{tunnel}+d\rho_t^{bath}+d\rho_t^{measure}$. The evolution through the tunnel effect is Hamiltonian: $d\rho_t^{tunnel}=-i[H,\rho_t]dt$. The dissipative evolution induced by the coupling to the left and right baths is modelled by a Lindblad equation: $d\rho_t^{bath}=\big( L_{\sigma^{+}_{l}}+L_{\sigma^{-}_{l}}+L_{\sigma^{+}_{r}}+L_{\sigma^{-}_{r}}\big) (\rho_t) \, dt$. The associated Lindblad generators $L_{\sigma^{+}_{l}}$, $L_{\sigma^{-}_{l}}$, $L_{\sigma^{+}_{r}}$, $L_{\sigma^{-}_{r}}$ take the form $L_\sigma(\rho)=\sigma\rho \sigma^\dagger - \frac{1}{2}\lbrace \sigma^\dagger \sigma, \rho\rbrace $ with $\sigma^{+}_{l}=\sqrt{a} e^{\beta \mu_l/2} \ket{L} \bra{0}$, $\sigma^{-}_{l} = \sqrt{a} \ket{0} \bra{L}$, $\sigma^{+}_{r}= \sqrt{b} e^{\beta\mu_r/2} \ket{R} \bra{0}$ and $\sigma^{-}_{r} = \sqrt{b} \ket{0} \bra{R}$. In this expression, $a$ (resp. $b$) is simply the coupling strength with the left (resp. right) electron reservoir and 
the general form provided insures that the system indeed converges to Gibbs state in the absence of tunnel coupling and measurement. The evolution due to the dot measurement back-reaction is: $d\rho_t^{measure}= L_{\mathcal{O}}(\rho_t) \, dt + D_{\mathcal{O}} (\rho_t) dW_t$, where ${\mathcal O}=h_l\ket{L}\bra{L}+h_r \ket{R}\bra{R}$ and $D_{\mathcal{O}} (\rho)= \left\lbrace \mathcal{O},\rho \right\rbrace-2\, \rho \, \mathrm{tr}(\mathcal{O}\rho)$, and $W_t$ a standard Wiener process, $(dW_t)^2=dt$, reflecting the random output of the weak indirect measurements. See Refs.\cite{wiseman2010A,barchielli1986A,barchielli1991A,barchielli2009A,belavkin1992A} for details.
Gathering:
\begin{equation}
\label{eq:first}
d\rho_t=\left[ -i[H,\rho_t] + \big[(L_{\sigma^{+}_{l}}+L_{\sigma^{-}_{l}})+(L_{\sigma^{+}_{r}}+L_{\sigma^{-}_{r}})+ L_{\mathcal{O}}\big](\rho_t)\right] \:dt 
+ D_{\mathcal{O}} (\rho_t) dW_t
\end{equation}

With those notations, and parametrizing $\rho_t$ as in the text, we get coupled stochastic differential equations (SDEs) for the density matrix components (with It\^o convention):
\begin{equation}\left\{
 \begin{split}
  dQ_0&=\big(a Q_l+b Q_r-(a m_l+b m_r) Q_0\big)\, dt -2 \big(h_l Q_l + h_r Q_r\big) Q_0\, dW_t, \\
  dQ_l&=\big(-uK+a m_l Q_0-a Q_l\big) \, dt  +  2\left(h_l(1-Q_l )-h_r Q_r\right) Q_l \,dW_t , \\
  dQ_r&=\big(uK-b Q_r+b m_r Q_0\big) \, dt + 2\left(h_r(1-Q_r)-h_l Q_l\right) Q_r \, dW_t ,\\
  dK&=2\big(-\nu_h\, K + u(Q_l-Q_r)\big) \, dt + (h_l(1-2Q_l)+h_r(1-2Q_r))K \: dW_t,
 \end{split} \right.
\end{equation}
where we used the simplified notations $m_l=e^{\beta \mu_l}$, $m_r=e^{\beta \mu_r}$ and $\nu_h= \left((h_l-h_r)^2 +a+b\right)/4$. We further take $h_r=-h_l=h$ to simplify the equations and make the assumption that the measurement rate is much bigger than the characteristic coupling rate with the baths, i.e.  $h^2\gg a\,\mathrm{and}\, b$. Then $\nu_h\simeq h^2$. Note that these equations are still redundant as the conservation of probability, i.e. the fact that $\mathrm{tr}(\rho)=1$, yields $Q_0+Q_l+Q_r=1$. Positivity of the DQD density matrix imposes $0\leq Q_{0,l,r}\leq 1$ and $K^2\leq 4Q_lQ_r$. The stochastic flow generated by (\ref{eq:Expanded}) takes place on the simplex $0\leq Q_{0,l,r}\leq 1$ and $K^2\leq 4Q_lQ_r$ with $Q_0+Q_r+Q_l=1$. We  denote by $0$, $L$ and $R$ the three corners of this simplex with coordinates $(Q_0=1, Q_l=0,Q_r=0)$, $(Q_0=0, Q_l=1,Q_r=0)$ and $(Q_0=0, Q_l=0,Q_r=1)$, respectively.

Different mechanisms are in competition in eq.(\ref{eq:Expanded}), and the typical behaviour of the DQD quantum trajectories results from this competition. The Hamiltonian flow induces a rotation on $K$ and $Q_{l,r}$ at a frequency of order $u$ and the Lindladian terms $L_{\sigma^\pm_{r,l}}$ a relaxation the DQD matrix towards the diagonal Gibbs density matrix. The $h$-dependent terms describe the random progressive collapses on eigen-vectors of ${\cal O}$ due to the measurement. They thus force the DQD density matrix to converge to one of the corners of the simplex. The characteristic time for collapsing is of order $h^{-2}$, and therefore very small in the limit of interest. There is thus a separation of time scales, the shortest one being that associated to the measurement collapse, and this is the origin of the jumpy behaviour of the DQD trajectories.

\begin{figure}
 \includegraphics[width=\textwidth]{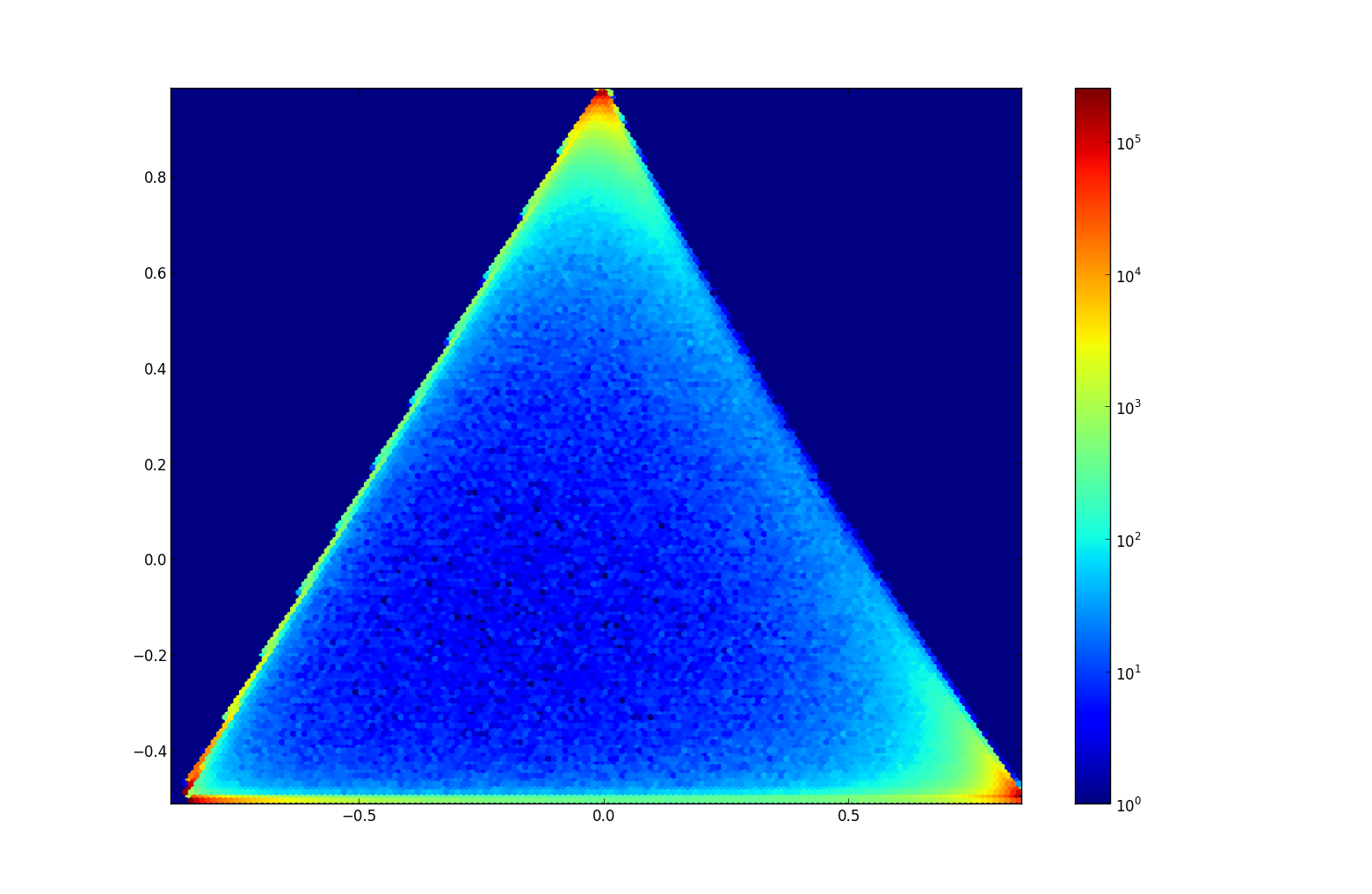}
 \caption{Invariant measure computed with the Monte-Carlo method for $h_l=-h_r=7.0$, $u=1.25$, $\mu_l=\mu_r=0$, $a=b=0.02$. The lower left vertex is the state $\ket{0}$ and the upper corner $\ket{R}$.}
 \label{fig:pdf}
\end{figure}

Echoes of these competitive mechanisms can be seen in the invariant measure of the process (\ref{eq:Expanded}). As can be seen in  Fig.\ref{fig:pdf}, it is highly concentrated on the corners $0, R,L$ of the simplex $0\leq Q_{0,l,r}\leq 1$. By ergodicity this means that a typical trajectory spends a large amount of time close to the corners, and abruptly jumps at random times from one corner to the other. The time duration of these jumps are of order $h^{-2}$ up to logarithmic corrections. There are two types of transitions, either between $0$ and $R$ or $L$, or between $R$ and $L$, which are generated by two different mechanisms: the transitions $0R$ or $0L$ are generated by the dissipative Lindbladian evolution, while the transitions $RL$ or $LR$ are Hamiltonian generated.

We now give estimations of the transition rates for the transitions from one corner to another which yield to the approximate $3$-state Markov chain description of the DQD quantum trajectories. The states of the Markov chain are in one-to-one correspondence to the corners of the simplex, so that we safely index them as $0$, $R$ or $L$.

\subsection{A reductive approach to the transition rates.}
Here we give an estimate (at $h$ large) of the transition rates assuming we can deal with the different transition processes independently.

\underline{\it $0\leftrightarrow L$ (or $0\leftrightarrow R$) transition}: As suggested by the numerics, and in particular by the shape of the invariant measure, let us assume that the $0\leftrightarrow L$ transition process is dominated by trajectories close to the $0L$ boundary of the simplex for which $Q_r\simeq 0$ (and hence $K\simeq0$ and $Q_l \simeq 1-Q_0$). Under these approximations, eq.(\ref{eq:Expanded}) then becomes
\begin{equation}
dQ_0=\lambda (p-Q_0) dt +2 h (1-Q_0) Q_0\, dW_t, 
\end{equation}
with $\lambda=a(m_l+1)+b m_r$ and $p=a/\lambda$. This evolution codes for the competition between the bath coupling and the measurement back-action, and it was studied in full details in \cite{thermalImagingA}. The sharp transitions emerge from an extension of Kramers theory where a particle excited only by white noise fluctuations shows a jumpy behaviour because it evolves in a bimodal effective potential\footnote{The process which is involved here is not exactly Kramer's like because quantum trajectories with strong measurement actually deal with large, but multiplicative and of a peculiar form, noises. Nevertheless the presence of a separation of time scales guarantees the Kramer's like behaviors.}. The jump statistics from $Q_0=0$ to $Q_0=1$ (or the reversed) becomes Poissonian in the large $h$ limit, the transition rates, defined as the mean time between two jumps, were shown to converge to $\lambda_{0L}=am_l$, $\lambda_{L0}=a$ (and similarly $\lambda_{0R}=bm_r$, $\lambda_{R0}=b$) for $h$ large. Notice 
that the transition rates converge to a finite limit when the measurement strength goes to infinity\footnote{That is: there is no QZE for thermally activated quantum jumps. This is of course only the case as long as we are still in the dissipative regime. If the measurement is intense enough to freeze all the degrees of freedom of the bath itself, then we will indeed see QZE.}. 

\underline{\it $L\leftrightarrow R$ transition}:
Similarly, we assume that the trajectories involved in this transition are predominantly close to the $LR$-boundary, i.e. $Q_0\simeq 0$ (and hence $Q_l\simeq 1-Q_r$), so that the evolution is dominated by the competition between the measurement back-action and Hamiltonian evolution. With those simplifications in mind, the set of equations (\ref{eq:Expanded}) simplifies to:
\begin{equation}
\begin{split}
dQ_r&=(uK -bQ_r)\, dt + 4h (1-Q_r)\, Q_r \, dW_t \\
dK&=2\left( -\nu_h K + u(1-2Q_r)\right) \, dt  +  2h\left(1-2Q_r \right) K \: dW_t
\end{split}
\end{equation}
These equations, but in the limit where $bQ_r\ll uK$ and for $h\gg a$ and $b$ so that $\nu_h\simeq h^2$, describe continuous measurement of an observable non-commuting with the Hamiltonian evolution of a two level system. They of course induce quantum jumps between the observable eigen-states, and were studied with mathematical details in \cite{qbmA}.  For large $h$, the jump statistics becomes Poissonnian and the average time between two jumps gives a transition rate $\lambda_{LR} = \lambda_{RL} = u^2/h^2$. Notice that this rate goes to 0 when the measurement strength goes to infinity as expected from the quantum Zeno effect (QZE).

\subsection{Decomposition of the DQD evolution and the transition rates.}
We here describe an alternative decomposition of the dynamics of DQD quantum trajectories in two intrinsically different processes which take places on different time scales: those associated to the behaviour when the DQD trajectory is close to one the corner, and those associated to the jumps from one corner to the other. The later are dominated by the measurement process (for $h$ large), and hence by the collapse of the DQD density matrix. This decomposition yields a finer estimate of the probabilities to transit from one corner to another and to the transition rates. 

\underline{\it Transition to $L$ or $R$ starting from $0$}:
We are here aiming at evaluating the probability $\Pi_{0\to L}$ (or $\Pi_{0\to R}$) to reach the $L$-corner first (or the $R$-corner first) starting from $0$. These are the transition probabilities starting from $0$, and of course $\Pi_{0\to L}+\Pi_{0\to R}=1$. 
 
Let us slice the simplex $0\leq Q_{0,r,l}\leq 1$ by cutting the $0$-corner with a surface at an intermediate scale, defined say by $Q_0=1-\epsilon$, with $\epsilon \ll 1$ (or equivalently by $Q_r+Q_l=\epsilon$). Any DQD quantum trajectory starting at $0$ will eventually hit this surface at a random position. Actually it will hit this surface an infinite number of times. To be quantitative, let us look at the probability distribution\footnote{Mathematically, this distribution is called the {\it harmonic measure} of the process starting at $0$.} of the position at which the trajectory starting at $0$ hit the surface $Q_0=1-\epsilon$ for the first instance.
Let us denote averages with respect to this probability distribution by $\langle\cdots \rangle_\epsilon$. For $h$ large, the dominating process acting on the DQD 
trajectory in the neighbourhood of a surface $Q_0=1-\epsilon$ at an intermediate scale is the measurement process. Thus, once the DQD trajectory has reached the surface $Q_0=1-\epsilon$, it collapses to one of the corners, either back at the $0$-corner or towards the $R$- or $L$-corner, and this collapse takes place on a microscopic time scale of order $h^{-2}$. The probability $\hat \Pi^\epsilon_{0,R,L}$ for collapsing to either one of the corners are given by the Von Neumann rules of quantum measurement\footnote{As explained above, these rules are encoded in the stochastic differential equations (\ref{eq:Expanded}).}, so that $\hat \Pi^\epsilon_{0,R,L}=\langle Q_{0,r,l}\rangle_\epsilon$ because on the intermediate surface the initial position of the DQD trajectory are distributed according to the measure $\langle\cdots\rangle_\epsilon$\footnote{Here, we implicitly assume that the successive hitting positions of the DQD trajectory when hitting again the surface after having collapse back towards $0$ are 
uncorrelated. This is justified because the elementary processes which are involved here are those associated to the collapse and they take place on very small time scales of order $h^{-2}$. }. Taking into account that after having hit the intermediate surface the trajectory may go back to $0$ an arbitrary number of times (these correspond to interrupted attempts to transit from $0$ to $L$ or $R$), we get for the transition probabilities as
\begin{equation}
\Pi_{0\to L} = \frac{\hat \Pi^\epsilon_{L} }{1-\hat \Pi^\epsilon_{0} }=\frac{\langle Q_l\rangle_\epsilon }{1-\langle Q_0\rangle_\epsilon},\quad 
\Pi_{0\to R} = \frac{\hat \Pi^\epsilon_{R}}{1-\hat \Pi^\epsilon_{0}}=\frac{\langle Q_r\rangle_\epsilon }{1-\langle Q_0\rangle_\epsilon}. 
\end{equation}
Recall that $Q_0+Q_r+Q_l=1$ and hence $\Pi_{0\to L}+\Pi_{0\to R}=1$.
We now need an estimation of the mean hitting position $\langle Q_{0,r,l}\rangle_\epsilon$. To get it, we pick $\epsilon$ small enough so that the noisy terms in eq.(\ref{eq:Expanded}) become sub-dominant and the DQD trajectories are then governed (approximatively, but more and more exactly as $\epsilon$ is decreased) by the deterministic drift terms. For $Q_{r,l}$ small, that is to leading order in $\epsilon$, the SDEs (\ref{eq:Expanded}) become $dQ_l\simeq am_l\,dt+\cdots$ and $dQ_r\simeq  bm_r\, dt+\cdots$, so that $\langle Q_l\rangle_\epsilon \simeq \epsilon\, am_l/(am_l+bm_r)$ and $\langle Q_r\rangle_\epsilon \simeq \epsilon\, bm_l/(am_l+bm_r)$. Hence,
\begin{equation}
\Pi_{0\to L} \simeq \frac{am_l}{am_l+bm_r},\quad 
\Pi_{0\to R} \simeq \frac{bm_r}{am_l+bm_r.} 
\end{equation}
We actually get a bit more because we can also estimate the mean time $\tau_0$ the DQD trajectories stay near the $0$-corner. Indeed, within the above approximation scheme, the typical time to reach for the first instance the intermediate surface starting from $0$ is of order $\epsilon/(am_l+bm_r)$. Taking again into account the fact that the DQD trajectories may go back to $0$ an arbitrary number of times after having hit the intermediate surface, we get
\begin{equation}
\tau_0\simeq (\frac{\epsilon}{am_l+bm_r})(\frac{1}{1-\hat \Pi^\epsilon_{0}})\simeq \frac{1}{am_l+bm_r}.
\end{equation}

\underline{\it Transition to $0$ or $R$ starting from $L$ (or to $0$ or $L$ starting from $R$):}
We here implement the same strategy by slicing the simplex at an intermediate scale $\epsilon$ away from the $L$-corner. We similarly get that the transition probabilities starting from $L$ are
\begin{equation}
\Pi_{L\to 0} = \frac{\langle Q_0\rangle_\epsilon }{1-\langle Q_l\rangle_\epsilon},\quad 
\Pi_{L\to R} = \frac{\langle Q_r\rangle_\epsilon}{1-\langle Q_l\rangle_\epsilon}, 
\end{equation}
where $\langle\cdots\rangle_\epsilon$ now denote averages with respect to the probability distribution of the hitting position on the intermediate surface at a distance $\epsilon$ from $L$. We still have to evaluate the mean hitting positions $\langle Q_{0,l,r}\rangle_\epsilon$, the only difference compared to the previous discussion is that now the $K$-dynamics is relevant. As above we estimate this mean position by linearizing the dynamics for $Q_0,\, Q_r$ small. To leading order, we have $dQ_0=a\, dt+\cdots$, $dQ_r=uK\, dt+\cdots$ and $dK=2(-\nu_hK+u)\, dt +\cdots$. Hence, asymptotically $K\simeq u/\nu_h$ and the mean hitting position is at $\langle Q_0\rangle_\epsilon\simeq \epsilon\, a/(a+ u^2/\nu_h)$ and $\langle Q_r\rangle_\epsilon\simeq \epsilon\, (u^2/\nu_h)/(a+ u^2/\nu_h)$. Thus
\begin{equation}
\Pi_{L\to 0} \simeq \frac{u^2/\nu_h}{a+ u^2/\nu_h},\quad \Pi_{L\to R} \simeq \frac{a}{a+ u^2/\nu_h}.
\end{equation}
The mean time $\tau_L$ the QDQ trajectories spend near the $L$-corner is evaluated as above:
\begin{equation}
\tau_L \simeq (\frac{\epsilon}{a+ u^2/\nu_h})(\frac{1}{1-\hat \Pi^\epsilon_{L}})\simeq \frac{1}{a+ u^2/\nu_h}.
\end{equation}
The transition rates starting from the $R$-corner are evaluated similarly by exchanging $a$ and $b$.

\subsection{Markov chain limit.} 
All formulas for  $\Pi_{0\to L,R}$ or $\Pi_{R,L\to 0}$ or $\Pi_{R\leftrightarrow L}$, as well as for the mean times spend around each of the corners, correspond to the transition probabilities for a Markov chain with transition rates 
\begin{equation}
\lambda_{0R}=am_l,\ \lambda_{0L}=bm_r,\ \lambda_{R0}=a,\ \lambda_{L0}=b,\ \lambda_{LR}=\lambda_{RL}=u^2/\nu_h.
\end{equation}
We thus conclude that the DQD trajectory may approximatively be described by an effective $3$-state Markov chain whose master equation reads
\begin{equation}
\left\{
\begin{split}
\partial_t \Pi_0&=\left(-a m_l-bm_r\right) \Pi_0 + b \Pi_r + a \Pi_l\\
\partial_t \Pi_l&=-(a+\frac{u^2}{\nu_h})\Pi_l +\frac{u^2}{\nu_h} \Pi_r +am_l \Pi_0 \\
\partial_t \Pi_r&=-(b+\frac{u^2}{\nu_h})\Pi_r + \frac{u^2}{\nu_h} \Pi_l +bm_r \Pi_0
\end{split}  \right.
\label{eq:masterBis}
\end{equation}
with $\Pi_{0,r,l}$ the probabilities for the DQD trajectory to be in the state $0$, $R$ or $L$; that is, $\Pi_{0,l,r}$ are the probabilities for the DQD quantum trajectory to be near the $0$-, $R$-, or $L$-corners.
Notice that equations (\ref{eq:Expanded}) for the DQD quantum trajectories do not contain any counting process. For a strong measurement  intensity, the white noise excitation added to the non linearity of the equations are enough to give rise to effectively punctual jumps.

\subsection{Quantum mechanical and statistical evaluation of the flux.}
Here we prove that the mean flux computed ``quantum mechanically" using the mean steady DQD density matrix\footnote{Here, taking the mean, with respect to the Wiener process, amounts not to record the results of the dot measurements but averaging over them.} coincides with the mean ``statistical flux" computed using the effective $3$-state Markov chain. Let $\bar Q_{0,r,l}^{steady}$ and $\bar K^{steady}$ be the component of the mean steady DQD density matrix $\bar \rho^{steady}$. By definition they satisfy:
\begin{equation}\left\{
 \begin{split}
  a\bar Q_l^{steady}+b \bar Q_r^{steady}-(a m_l+b m_r) \bar Q_0^{steady}=0\\
  -u\bar K^{steady}+a m_l \bar Q_0^{steady}-a \bar Q_l^{steady}=0 \\
u\bar K^{steady}-b \bar Q_r^{steady}+b m_r \bar Q_0^{steady}=0 \\
-\nu_h\, \bar K^{steady} + u(\bar Q_l^{steady}-\bar Q_r^{steady})=0  \end{split} \right.
 \label{eq:MeanSteady}
\end{equation} 
with $\bar Q_0^{steady}+\bar Q_r^{steady}+\bar Q_l^{steady}=1$. The quantum mechanical mean flux is $\langle J \rangle_{MQ}= {\rm tr}( \hat J\, \bar \rho^{steady})$ with $\hat J= iu(|R\rangle\langle L|-|L\rangle\langle R|)$, and hence
\begin{equation}
\langle J \rangle_{MQ}= u \bar K^{steady} = (u^2/\nu_h)\, (\bar Q_l^{steady}-\bar Q_r^{steady}).
\end{equation}
Note that this mean flux is computed using the steady DQD density matrix with the measurement back-action taken into account.
On the other hand, the stationary measure $\Pi^{stat}_{0,r,l}$ of the effective Markov chain satisfies:
\begin{equation} 
\left\{ \begin{split}
\left(-a m_l-bm_r\right) \Pi_0^{stat} + b \Pi_r^{stat} + a \Pi_l^{stat}=0\\
-(a+\frac{u^2}{\nu_h})\Pi_l^{stat} +\frac{u^2}{\nu_h} \Pi_r^{stat} +am_l \Pi_0^{stat}=0 \\
-(b+\frac{u^2}{\nu_h})\Pi_r^{stat} + \frac{u^2}{\nu_h} \Pi_l^{stat} +bm_r \Pi_0^{stat}=0.
\end{split} \right.
\label{eq:MarkovStat}
\end{equation}
with $\Pi_0^{stat}+\Pi_r^{stat}+\Pi_l^{stat}=1$. The mean statistical flux, defined as the difference between the numbers of $LR$ and $RL$ transitions per unit of time is  $\langle J\rangle_{stat}=\lambda_{LR}\Pi^{stat}_l-\lambda_{RL}\Pi^{stat}_r$ by standard arguments\cite{jacod2003A}. Hence
\begin{equation}
\langle J\rangle_{stat}=(u^2/\nu_h)\, (\Pi^{stat}_l-\Pi^{stat}_r).
\end{equation}
Using $u\, \bar K^{steady} = (u^2/\nu_h)\, (\bar Q_l^{steady}-\bar Q_r^{steady})$, one readily sees that $\Pi_{0,r,l}^{stat}=\bar Q_{0,r,l}^{steady},$ so that
\begin{equation}
\langle J\rangle_{stat}=\langle J \rangle_{MQ}.
\end{equation}
To be complete, let us give the expression of the mean stationary DQD state:
\begin{equation}
\left \{ \begin{split}
\bar Q^{steady}_0&= \big[ab\, \nu_h+ (a+b)\, u^2\big]/\Delta \\
\bar Q^{steady}_l&= \big[abm_l\, \nu_h+ (am_l+bm_r)\, u^2\big]/\Delta \\
\bar Q^{steady}_r&= \big[abm_r\, \nu_h + (am_l+bm_r)\, u^2\big]/\Delta 
\end{split} \right.
\label{eq:masterTer} 
\end{equation}
with $\Delta= ab(m_l+m_r+1)\, \nu_h+ (a(2m_l+1)+b(2m_r+1))\, u^2$.
Fluctuations of charge transfer can be approximatively computed using large deviation techniques for Markov chains. Standard results of probability theory therefore apply, and in particular the fluctuation relations\cite{gallavotti1995A,cohen1999A} apply to the charge transfer statistics at least within the Markov chain approximation.

\section{Numerics}
Computing numerically the average flux  directly from the stochastic differential equation is an easy but cumbersome task. For a finite measurement strength, the system will not be exactly in the basis state vectors, i.e. the quantum trajectory will not remain exactly on the vertices of the triangle we previously defined. Therefore, one needs to define a finite radius around the edges in order to be able count the number $N_{LR}$ of transitions from $\ket{L}$ to $\ket{R}$ (resp. $N_{RL}$ from $\ket{R}$ to $\ket{L}$) and compute the flux. Fortunately, the results do not depend on this cutoff at the strong measurement limit. It is possible to count those transitions in many other ways, for example by counting the number of turns the quantum trajectory makes around the centre of the triangle. Once again, the numerical results show no significant differences between those different methods.

The main problem is thus not conceptual but practical. To get accurate results, one needs to record many jumps that get rarer as the measurement intensity increases. A strong measurement intensity also means that the jumps get sharper thus requiring a small discretisation step. As a consequence, naively discretising the SDE using a Euler scheme and a constant step proves to be insufficient to compute the flux in a reasonable amount of time. We first use a second order Runge-Kutta discretisation scheme adapted to SDE\cite{kloeden1977A} that strongly increases the computation speed. For a given time step $dt$ we compute the elementary evolution in the following way:
\begin{equation}
\begin{split}
 &\rho_{t+dt}=\rho_t +  \mathcal{L}(\rho_t)\: dt  +  D_{\mathcal{O}}(\rho_t) \:\epsilon_t +  \frac{1}{2}\big(D_{O}(\tilde{\rho}_t)-D_{O}(\rho_t)\big) (\epsilon_t^2-dt) \: dt^{-1/2} \\
 &\mathrm{with } \;\; \tilde{\rho}_t = \rho_t +  \mathcal{L}(\rho_t) \: dt + D_{\mathcal{O}}(\rho_t)\: \sqrt{dt} 
\end{split}
\end{equation}
Where we have collected the deterministic part of eq.(\ref{eq:first}) in the superoperator $\mathcal{L}$ and written $\epsilon_t$ a random number drawn from a normal distribution $\mathcal{N}(0,\sqrt{dt})$.

The situation can be further improved using an adaptive discretisation scheme. Indeed, the system spends most of the time evolving slowly around basis state and very rarely jumps in a sharp evolution. A constant step discretisation scheme is thus unnecessarily precise most of the time and not precise enough during the jumps. We have used a simple algorithm that refines the discretisation during the jumps.

\begin{figure}
 \includegraphics[width=\textwidth]{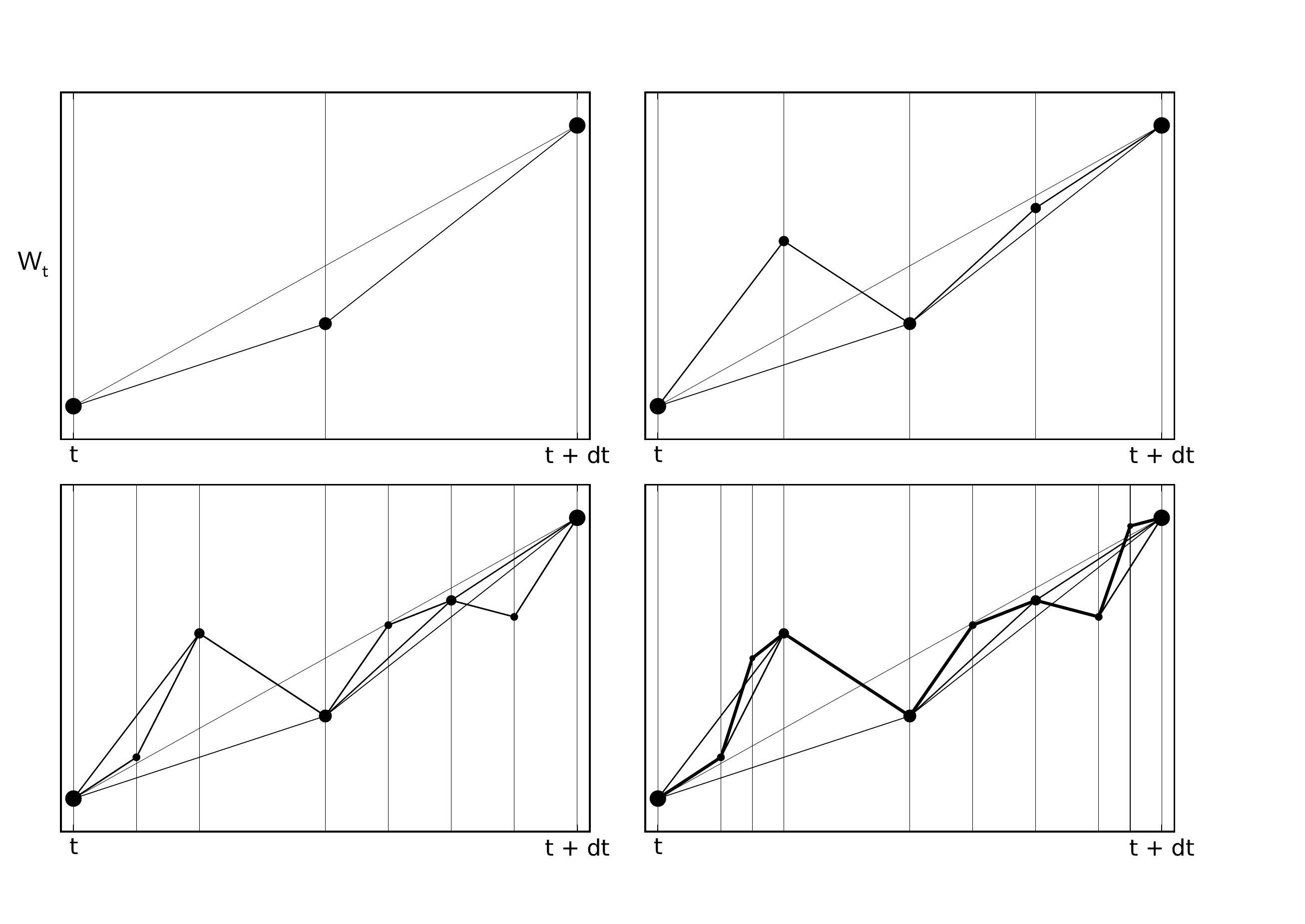}
 \caption{Schematic view of the non-uniform refining procedure of $W_t$ starting from a coarse grained sampling. }
 \label{fig:refining}
\end{figure}

We first start with a coarse grained time step $dt$ to discretise the noise. At time $t$, we compute $\rho_{t+dt}$ from $\rho_t$ -- using Runge-Kutta algorithm -- which requires the drawing of one random number $\epsilon$ from a normal distribution $\mathcal{N}(0,\sqrt{dt})$. We use Paul L\'evy's construction of Brownian motion (see \cite{levy1965A,hida1980A}) which is based on dichotomy and interpolation to refine the sampling, i.e. we draw an intermediary random number $\epsilon_{1/2}$ from a normal distribution $\mathcal{N}(\epsilon/2,\sqrt{dt/2})$. We then compute $\rho_{t+dt}^{refined}$ from this finer discretisation and compute the error cost function $tr \left[ \left(\rho_{t+dt}^{refined}-\rho_{t+dt}\right)^2 \right]$. If the cost is bigger than a given error threshold, we apply the same procedure as before on the two intervals of size $dt/2$ and keep on recursively as long as needed (see Fig. \ref{fig:refining}). The important point to notice is that we use every single random number that is drawn so 
that the adaptive scheme does not bias the evolution. 

In our computations, we have taken a small enough error threshold to insure that the discretisation error remains much smaller than the pure Monte-Carlo error that arises from the finite number of jumps recorded. The error bars provided in this paper are thus pure Monte-Carlo errors. As the flux computation, i.e. jump counting, is a trivially parallelisable problem, the use of a computing grid is an other less clever yet dramatically efficient way to improve the computation speed and reduce the Monte-Carlo errors.

\newpage

\end{document}